\documentclass[12pt]{article}
\usepackage{graphicx}
\begin{document}
\rightline{NKU-2014-SF2}
\bigskip

\newcommand{\be}{\begin{equation}}
\newcommand{\ee}{\end{equation}}
\newcommand{\noi}{\noindent}
\newcommand{\refb}[1]{(\ref{#1})}
\newcommand{\ra}{\rightarrow}
\newcommand{\bib}{\bibitem}

\begin{center}
{\Large\bf  Null trajectories and bending of light in charged black holes with  quintessence}

\end{center}
\hspace{0.4cm}
\begin{center}
Sharmanthie Fernando \footnote{fernando@nku.edu}, Scott Meadows \footnote{meadowss3@mymail.nku.edu}, and 
Kevon Reis \footnote{reisk3@mymail.nku.edu}\\
{\small\it Department of Physics \& Geology}\\
{\small\it Northern Kentucky University}\\
{\small\it Highland Heights}\\
{\small\it Kentucky 41099}\\
{\small\it U.S.A.}\\

\end{center}

\begin{center}
{\bf Abstract}
\end{center}

\hspace{0.7cm} 

We have studied   null geodesics of  the charged black hole surrounded by  quintessence.  Quintessence is a candidate for dark energy and is represented by a scalar field. Here, we have done a detailed study of the photon trajectories. The exact solutions for the trajectories are obtained in terms of the Jacobi-elliptic integrals  for all possible energy and angular momentum of the photons.  We have also studied the bending angle using the Rindler and  Ishak method.\\

{\it Key words}: static, charged, quintessence, null geodesics, deflection of light


\section{ Introduction}

Recent observations support that the universe is undergoing expansion at an accelerated rate \cite{perl}\cite{reis}. The unknown cause for this cosmic acceleration is known as ``dark energy'' which consists of about $70\%$ of the energy density of the universe. Understanding the nature of this unknown energy component  in the universe is one of the greatest challenges in modern cosmology. An interesting review related to dark energy  is given in \cite{pee}. A thermodynamic motivation for dark energy can be found in \cite{rad}. Observational constraints on dark energy can be found in \cite{wang}.

The equation  of state parameter for dark energy is given by $\omega = \frac{p}{\rho}$. Here $p$ is the pressure and $\rho$ is the energy density of the dark energy. For acceleration to occur, the pressure has to  be negative. There are large number of surveys proposed to find the value of $\omega$ and its time evolution \cite{hea} \cite{ref}. .  

One of the  candidates for  dark energy density considered in literature is the cosmological constant with a state parameter $\omega = -1$.  There is a  major problem that is yet to be understood about the cosmological constant from a fundamental physics point of view. The observed value is too small in comparison with the theoretical prediction  and this  is well known as the fine-tuning problem \cite{wein1}. 

Current observations seems to be consistent with a value of $\omega = -1$. However, these observations say relatively little as to how $\omega$ evolve with time. Therefore, it is important to consider alternative models of dark energy where the equation of state changes with time, such as in inflationary  cosmology. So far, wide variety of dark energy models with dynamical scalar fields have been proposed as alternative models to cosmological constant. Such scalar field models include, but not limited to, quintessence, K-essence, ghost condensates, phantoms, and dilaton dark energy. A review of the quintessence can be found in \cite{edmund}
\cite{shin}.

In this paper we focus on quintessence as the candidate for dark energy.  Quintessence is described by a scalar field minimally coupled to gravity with the action,
\be \label{quin}
S_{quintessence} = \int d^4x \sqrt{-g}  \left( - \partial_{\mu} \Phi  \partial^{\mu} \Phi - V(\Phi) \right)
\ee
The energy momentum tensor for such a field is given by,
\be
T_{\mu \nu} = \partial_{\mu} \Phi \partial_{\nu}\Phi - g_{\mu \nu} \left( \frac{1}{2} \partial^{\sigma}\Phi \partial_{\sigma} \Phi + V(\Phi) \right)
\ee
For example, in a flat Friedmann-Robertson-Walker  universe, the expressions for pressure and energy density for the quintessence is given by,
\be
\rho_q = T^0_0 =  \frac{{\dot{\Phi}}^2}{2} + V(\Phi)
\ee
\be
p_q = T^1_1 = \frac{\dot{{\Phi}}^2}{2} - V(\Phi)
\ee
For an accelerated universe, $\frac{\dot{{\Phi}}^2}{2}  < V(\Phi)$. Hence for acceleration to occur, the field $\Phi$ should vary slowly along the potential $V(\Phi)$. Such a requirement is similar to the slow-roll inflation of the early universe. In addition to the above requirement, for acceleration to occur, the mass of the quintessence $m_{\Phi} = \sqrt{\frac{ d^2V}{d\Phi^2}}$ needs to be very small. This means $m_{\Phi} \leq H_0 \approx 10^{-33} eV$ where $H_0$ is the Hubble parameter today.

Many different potentials for the quintessence have been studied in the literature. Depending on how the state parameter evolve in these models, quintessence models are classified in two classes: One is the thawing model. In this case, the field freezes during the cosmological epoch and  restart to evolve once the mass of the field drops to a very small value. The second model is the freezing model. Here, the potential tends to be shallow at late times leading the field to slow down. For example, a freezing model can have  the potential,
\be
V(\Phi) = M^{ 4 + p} \Phi^{-p}
\ee
where $M$ and $p (>0)$ are constants. A graph plotted in Fig.1 clearly demonstrate the potential being shallow at late times, leading to acceleration. A review with explicit calculations for various potentials for the quintessence is found in \cite{shin}.

\begin{center}
\scalebox{.9}{\includegraphics{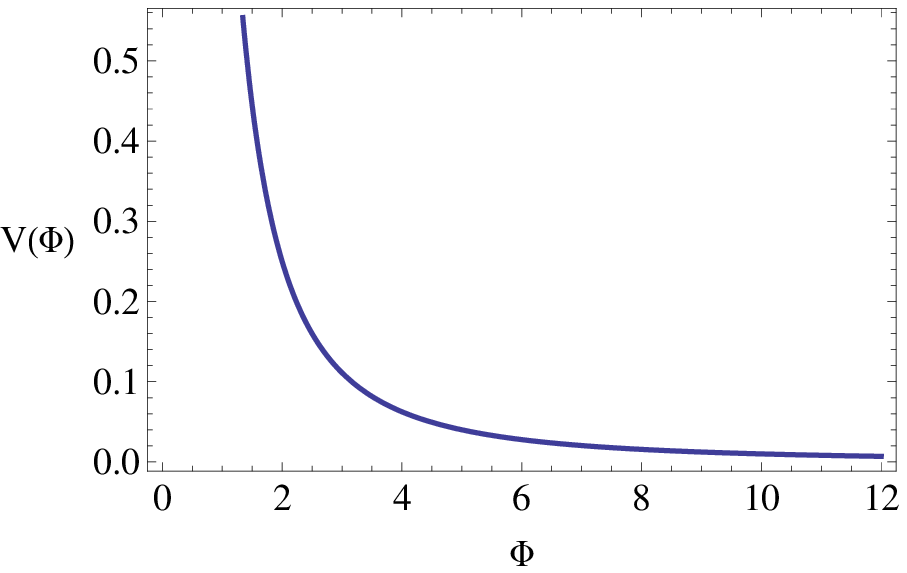}}

\vspace{0.3cm}

 \end{center}

Figure 1. The figure shows  $V(\Phi)$ vs $\Phi$ for the constants $M=1$, $p=2$\\

There are many works related to the quintessence model. We will present only a few here due to the large volume of existing work. Discussion about tracker solutions for conformally coupled quintessence model is discussed in \cite{shoja}. Another article on tracking quintessence is written by Lola et.al \cite{lola}. Phase space analysis of quintessence fields trapped in a Randall-Sundrum Braneworld is done by  
Escobar et.al. \cite{esco}. Laboratory search for quintessence model is given in \cite{romali}. A study on quintessence and phantom dark energy from ghost D-branes is presented by Saridakis and Ward in \cite{sar}.

The above discussion about the quintessence as a candidate for dark energy  is from a cosmological point of view.  Given the fact that black holes are a part of our cosmos, it is natural to ask how dark energy or quintessence in particular,  affects the formation, evolution and physics of black holes in general. In an interesting article, Li and Wang \cite{li} showed that black holes can exist in a Friedmann-Robertson-Walker universe dominated by dark energy. Ishwarchandra et.al. \cite{ishwa} produced an exact solution of a black hole in dark energy background with the state parameter $\omega = -\frac{1}{2}$. A detailed review of black holes in the presence of dark energy is given by Babichev et.al \cite{bab}. In this review wide range of field theoretical models were considered as dark energy.

In this paper we study how the quintessence effect the trajectories of massless particles around charged black holes. Studies of geodesics of massive and massless particles are one of the  ways to understand the gravitational field around a black hole. Theoretical  predictions related to geodesics such as gravitational lensing, perihelion shift, gravitational time delay and Lense-Thirring effect etc are aspects of black hole physics which can be compared to observations. From an astrophysical point of view, the study of orbits of test particles are important to understand  the flow of particles in accretion disks around black holes.  Furthermore, circular orbits of  photons (also known as ``photon sphere''\cite{vir}) are important in studying the structure of the black hole geometry \cite{hod}. Also, quasinormal modes of black holes are related to  null geodesics as shown in \cite{car}. Hence studying  null trajectories will help in understanding the stability properties of black holes to name a few applications. Considering all of above, we believe there is importance in doing a thorough understanding of trajectories in the presence of the dark energy element, quintessence.

The paper is organized as follows: In section 2, an introduction to charged black holes surrounded by the quintessence is given. In section 3,  null geodesics of the black hole is derived. In section 4,  radial null geodesics are studied. In section 5,  null geodesics with angular momentum and orbits of the photons corresponding to various values of the energy of photons  is presented in detail. In section 6, the bending of light is presented. Finally, in section 7 the conclusion is given.


\section { Charged  black hole surrounded by the quintessence}

In this section  we will present   the charged black hole surrounded by the quintessence studied in this paper. This particular black hole was derived by  Kiselev \cite{kiselev} and is describe by the metric,

\begin{equation}
ds^2 = - f(r) dt^2 + \frac{ dr^2}{ f(r)} + r^2 ( d \theta^2 + sin^2 \theta d \phi^2)
\end{equation}
where,
\begin{equation} \label{metric}
f(r) = 1 - \frac{ 2 M} { r} + \frac{ Q^2}{ r^2} - \frac{\alpha}{r^{ 3 \omega + 1}}
\end{equation}
Here, $M$ is the mass, $Q$ is the charge, and $\omega$ is the state parameter. $\alpha$ is a normalization factor. The energy density  and the pressure of the quintessence is given by,
\begin{equation} \label{rho}
\rho_q = - \frac{ \alpha}{ 2} \frac { 3 \omega}{ r^{ 3 ( 1 + \omega)}}
\end{equation}

\be
p_{q} = \omega \rho_q = - \frac{ \alpha  }{ 2} \frac { 3 \omega^2}{ r^{ 3 ( 1 + \omega)}}
\ee

For the range of $-1 < \omega < - \frac{1}{3}$, the universe with the quintessence will accelerate. General black hole solution for all $\omega$ were discussed  in \cite{kiselev}. In this paper, we pick $\omega = -\frac{2}{3}$ to facilitate computations. Hence, for this particular value of $\omega$, the  metric becomes
\begin{equation}
ds^2 = - f(r) dt^2 + \frac{ dr^2}{ f(r)} + r^2 ( d \theta^2 + sin^2 \theta d \phi^2)
\end{equation}
where
\begin{equation} \label{metric}
f(r) = 1 - \frac{ 2 M} { r} + \frac{ Q^2}{ r^2} - \alpha r
\end{equation}
As described by Kiselev \cite{kiselev}, the pressure of the quintessence matter has to be negative to cause acceleration. Therefore matter energy density $\rho_q$ is positive leading $ \alpha$  to be positive.

There are few works related to the   black hole described above. Thermodynamics of it has been studied  in \cite{sal} \cite{manu} \cite{fernando2}. Quasinormal modes of the charged black hole surrounded by the quintessence has been  studied in \cite{var}. Null geodesics of the Schwarzschild black hole surrounded by the quintessence was  studied by Fernando in \cite{fernando1}. The black hole solutions derived by Kiselev were extended to $d$ dimensions by Chen et.al. \cite{chen}. They also studied Hawking radiation of the $d$-dimensional black holes.


\subsection{Structure of horizons}

In order to study the trajectories, one has to understand where the horizons are located for the given black hole.
The horizons of the black hole is determined by the roots of the equation $ f(r)=0$, which leads to the cubic equation,
\begin{equation} \label{cubic}
\alpha r^3 - r^2 + 2 M r - Q^2 =0
\end{equation}
The nature of the roots of the polynomial in eq.$\refb{cubic}$ depends on the discriminant $\bigtriangleup$ given by,
\be \label{dis}
\bigtriangleup = 4 ( M^2 - Q^2) + \alpha( - 32 M^3 + 36 Q^2 M) - 27 \alpha^2 Q^4
\ee
When $ \bigtriangleup >0$, the polynomial will have three real roots. When $\bigtriangleup =0$, there will be two real roots (degenerate root and a real root). When $\bigtriangleup <0$ there will be only one real root. Hence, depending on the values of $M$, $Q$, and $\alpha$, the black hole could have three, two or one horizon.   A detailed discussion on how the roots behave when the parameters in the equation, $M$, $Q$ and $\alpha$  are changed, are given in the paper by Fernando \cite{fernando2}. In  Fig.2, a general class of plots for various values of the discriminant is given.

\begin{center}
\scalebox{.9}{\includegraphics{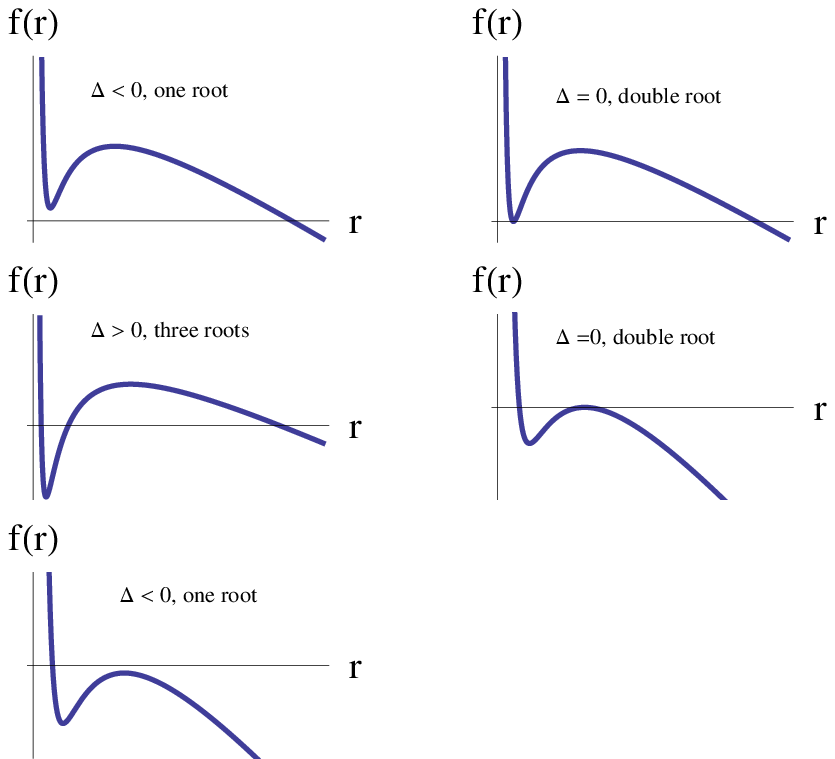}}

\vspace{0.3cm}

 \end{center}

Figure 2. The figure shows  $f(r)$ vs $r$ for the black hole with the quintessence.\\


\subsubsection{Three horizons}

When there are three roots for $f(r)$, the smallest root represents the black hole inner horizon ($r_+$). The second root represents the black hole event horizon ($r_{++}$). The largest root corresponds to the cosmological horizon $r_c$ which is similar to  what is observed in Schwarzschild-de Sitter black hole \cite{brum}. For small $\alpha$, $r_c \approx \frac{1}{\alpha}$ and $ M << \frac{1}{\alpha}$.  There is a static region between $r_{++}$ and $r_c$. In the rest of the paper, we will choose parameters where there are three horizons. Analytical expressions for the three horizons are given by,

\begin{equation}
r_c = 2 \sqrt{\frac{-\nu}{3}} cos\left( \frac{\psi}{3}\right) + \frac{ 1}{ 3 \alpha}
\end{equation}
\begin{equation}
r_{++} = 2 \sqrt{\frac{-\nu}{3}} cos\left( \frac{\psi}{3} - \frac{ 2 \pi}{3}\right) + \frac{ 1}{ 3 \alpha}
\end{equation}
\begin{equation}
r_{+} = 2 \sqrt{\frac{-\nu}{3}} cos\left( \frac{\psi}{3} - \frac{  \pi}{3}\right) + \frac{ 1}{ 3 \alpha}
\end{equation}
The three functions $\nu$, $\sigma$ and $\psi$ are given by,
\begin{equation}
\nu = \frac{ ( 6 M \alpha - 1)}{ 3 \alpha^2}
\end{equation}
\begin{equation}
\sigma = \frac{ (  - 2 + 18 \alpha M - 27 \alpha^2 Q^2)}{ 27 \alpha^3}
\end{equation}

\begin{equation}
\psi = cos^{-1}\left( \frac{ 3 \sigma}{ 2 \nu} \sqrt{ \frac{ - 3}{ \nu} }\right)
\end{equation}


\subsubsection{ Degenerate horizons}
When $\bigtriangleup =0$, and $ ( -1 + 6 M \alpha) \neq 0$, there are double roots, and a simple root for the function $f(r)$. The double root, given by $\rho$, and the simple root, given by $\sigma$, are given by the following

\begin{equation}
\rho = \frac{ (9 \alpha Q^2 - 2 M)}{ 2 ( -1 + 6 M \alpha)}
\end{equation}

\begin{equation}
\sigma = \frac{  (-1 + 8 \alpha M - 9 \alpha^2 Q^2)}{ \alpha ( -1 + 6 M \alpha)}
\end{equation}
The corresponding function $f(r)$ for the extreme black hole is,
\be
f(r) = \frac{ \alpha ( r - \rho)^2 ( \sigma - r)}{ r^2} 
\ee
When the function $\eta = -2 + 18 \alpha M - 27 \alpha^2 Q^2 >0$, the double root corresponds to $ r_{++}= r_c$ and the single root becomes $r_+$. Such black holes are called ``Nariai black holes.''  When $\eta = -2 + 18 \alpha M - 27 \alpha^2 Q^2  < 0$, the double root corresponds to $ r_+ = r_{++}$ and the simple root becomes $r_c$. Such black holes are called ``cold black holes.'' 
\be
r_+ = r_{++} = r_c = \frac{ 1}{ 3 \alpha} = \gamma
\ee
In this case, they  are  named ``ultra cold black holes" and the function $f(r)$ takes  the form,
\be
f(r) = \frac{- \alpha ( r - \gamma)^3}{r^2}
\ee

\section{ Null geodesics of the black hole}

In this section, we will derive  null geodesics  for the charged black hole with the quintessence. The method used here is similar to the approach given in the well known book by Chandrasekhar \cite{chandra}. The geodesics equations can be derived from the Lagrangian $\mathcal{L}$ given by

\begin{equation} \label{lag}
\mathcal{L}_{geo} = - \frac{1}{2} \left( - f(r) \left( \frac{ dt}{d\tau} \right)^2 +  \frac{1}{f(r)} \left( \frac{ dr}{d \tau} \right)^2 + r^2 \left( \frac{ d \theta} { d \tau} \right)^2 + r^2 sin^2 \theta \left( \frac{ d \phi} { d \tau} \right)^2 \right)
\end{equation}
where $\tau$ is an affine parameter along the geodesics. Due to the symmetries along $t$ and $\phi$ directions, there are two conserved quantities of the photons given by $E$ and $L$. These two quantities are related to $f(r)$  as,
\begin{equation} \label{tdot}
f(r) \dot{t} = E
\end{equation}
\begin{equation} \label{phidot}
r^2 sin^2 \theta \dot{\phi} = L
\end{equation}
In this paper the ``dot''  represents $\frac{d}{d \tau}$. We will consider the motion on the plane with $ \theta = \frac{ \pi}{2}$. Since the photon is confined to this plane, $\dot{\theta} =0$ and $ \ddot{\theta} =0$. With $ \dot{t}$ and $ \dot{\phi}$ given  by eq.$\refb{tdot}$ and eq.$\refb{phidot}$, the Lagrangian in eq.$\refb{lag}$ for photons becomes,
\begin{equation} \label{rdot}
\dot{r}^2 + f(r) \frac{L^2}{r^2} = E^2
\end{equation}
One can define an effective potential $V_{eff}= \frac{L^2 f(r)}{r^2}$ for the motion of the photon as,
\begin{equation} \label{newpot}
 \dot{r}^2 + V_{eff} = E^2
\end{equation}
By combining eq$\refb{newpot}$ and eq.$\refb{phidot}$,
\begin{equation} \label{drdphi}
\frac{ dr}{ d \phi} = \frac{r^2}{L} \sqrt{ E^2  - V_{eff}}
\end{equation}
The effective potential $V_{eff}$ can be expanded as,
\begin{equation}
V_{eff} = L^2 \frac{f(r)}{r^2} = \frac{ L^2}{r^2} - \frac{ 2 M L^2}{r^3} + \frac{Q^2 L^2}{r^4} - \frac{\alpha L^2}{r}
\end{equation}
The first term represents the centrifugal potential. The second term corresponds to the relativistic correction due to general relativity. The third term is due to the fact that the black hole has electric charge. The last term is the one due to the quintessence scalar field around the charged black hole. One can see that it leads to an attractive term. Since the term due to the quintessence is negative,  the potential is smaller compared to the one without the quintessence field. The potential does have a maximum and does not have a minimum between the cosmological horizon and the black hole horizon. Hence, the photons do not have a stable circular orbits around the black hole. Since there is  a maximum, the photons can have a unstable circular orbit.  Various orbits for non-zero angular momentum will be discussed in section(5).

The effective potential is plotted in Fig.3. for various values of $\alpha$. For large $\alpha$, the maximum height is smaller.  The potential is positive between the horizons since  the zeros of the $V_{eff}$ coincide with the two horizons, $r_{++}$ and $r_c$. Therefore, as long as the two horizons are non-degenerate, the potential will be positive in the region considered. When the horizons degenerate, the  maximum of the potential will be zero. If we further increase $\alpha$ (keeping $M$ and $Q$ constant), then the potential will become negative for all $r$ values. In this paper we will only study the behavior of photons for non-degenerate horizons.

\begin{center}
\scalebox{.9}{\includegraphics{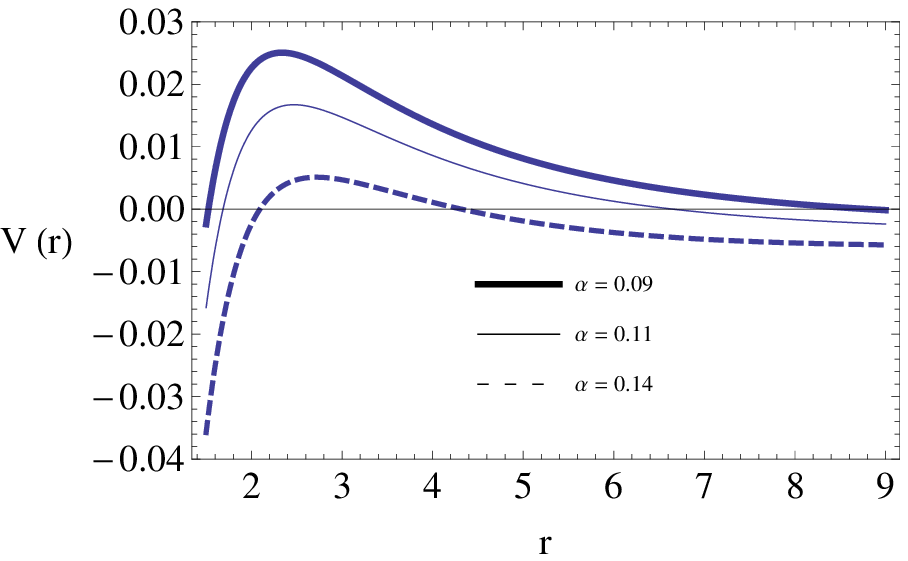}}

\vspace{0.3cm}

 \end{center}

Figure 3. The figure shows  $V_{eff}$ vs $r$ for the black hole with $M=0.96, L=1$ and $Q=0.96$.\\


\section{ Radial null geodesics}
First,  we like to study the trajectories of  photons without angular momentum. These trajectories lead to radial geodesics. This study is essential  to understand what happens to a freely falling photon towards the two horizons, $r_{++}$ and $r_c$.  First let us focus on the space-time for which $f(r)$ has three real roots leading to non-degenerate horizons.  For three real roots to exist, the discriminant given in eq.$\refb{dis}$ has to be greater than zero. 

If $f(r)$ has three real roots, then, $f(r)$ can be written as,
\begin{equation}
f(r) =  -\alpha \frac{( r - r_+) ( r - r_{++} ) ( r - r_c )}{r^2}
\end{equation}
For radial null geodesics, $ L=0$. Hence $ \dot{r}$ and $\dot{t}$ are given by,
\begin{equation} \label{dotrE}
\dot{r} = \pm E
\end{equation}
\begin{equation}
\dot{t} = \frac{E}{ f(r)}
\end{equation}
which leads to,
\begin{equation} \label{dtdr}
\frac{ dt}{ dr} = \pm \frac{1}{ f(r)} = \pm \frac{ 1 }{ ( 1 - \frac{ 2 M}{r} + \frac{ Q^2}{ r^2} - \alpha r ) }
\end{equation}
Before integrating  eq.$\refb{dtdr}$, some clarification is needed. The ``+'' sign is chosen for the the outgoing photons which reach the cosmological horizon. The ``-'' sign is chosen for the ingoing photons which reach the event horizon at $r_{++}$. Hence there will be two solutions for time $t$ as follows:

$$
t(event-horizon) = \frac{ 1}{\alpha ( r_+ - r_{++}) ( r_+ - r_c) ( r_{++} - r_c)} \left( r_+^2 ( r_{++} - r_c) Log( r - r_+) + \right. $$
\be
\left.  r_{++}^2 ( r_c - r_+) Log( r - r_{++} ) + r_c^2( r_+ - r_{++}) Log( r - r_c) \right) + const_{-}
\ee

$$
t(cosmological- horizon) = \frac{ -1}{\alpha ( r_+ - r_{++}) ( r_+ - r_c) ( r_{++} - r_c)} \left( r_+^2 ( r_{++} - r_c) Log( r - r_+) + \right. $$
\be 
\left.  r_{++}^2 ( r_c - r_+) Log( r - r_{++} ) + r_c^2( r_+ - r_{++}) Log( r - r_c) \right) + const_{+}
\ee

\noindent
The proper time can be obtained by integrating eq.$\refb{dotrE}$. Once again, ``-'' sign corresponds to the time towards the even horizon and ``+'' sign corresponds to the time towards the cosmological horizon.

\begin{equation} \label{proper}
\tau(event-horizon) =  -\frac{r}{E}  + const_{-}
\end{equation}

\begin{equation}
\tau(cosmological-horizon) =  \frac{r}{E}  + const_{+}
\end{equation}
\noindent
When $r \ra r_{++}, r_c$,  time $t$  goes to infinity as  shown in  Fig.4. The proper time is shown in Fig.5. and it is clear that it is finite when $r \rightarrow r_{++}, r_c$.
It is interesting to observe the contrasting  behavior of $t$ and $\tau$.  What this implies physically is that for an observer stationary at a point in between the two horizons, a photon radially falling towards the horizons will take an infinite time to reach them. On the other hand, the photons, in its own proper time will reach the horizons at finite time.

\begin{center}
\scalebox{.9}{\includegraphics{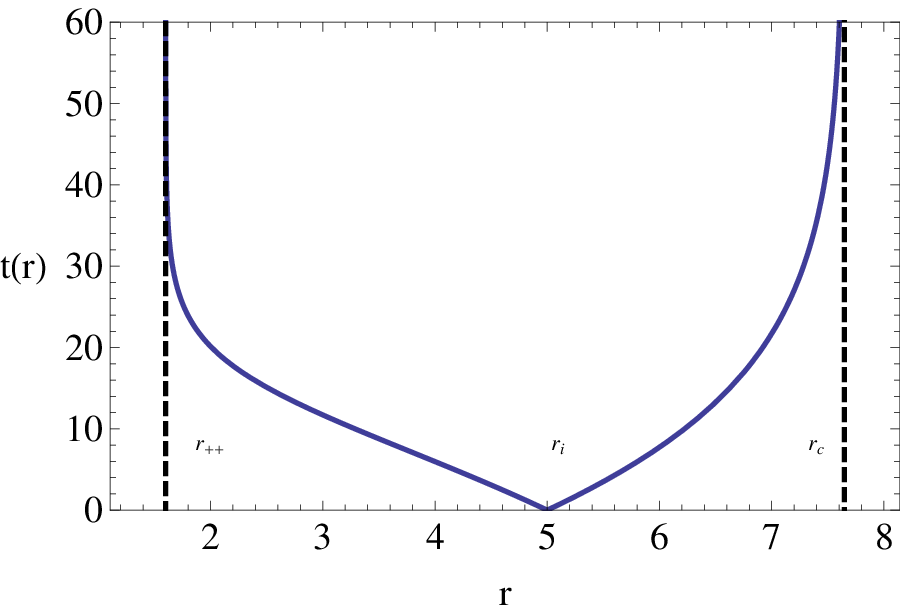}}
\vspace{0.3cm}
 \end{center}

Figure 4. The figure shows  $t$ vs $r$ for the black hole with $M=0.96$, $Q=0.96$. Here, $r =5$ at $ t =0$.\\

\begin{center}
\scalebox{.9}{\includegraphics{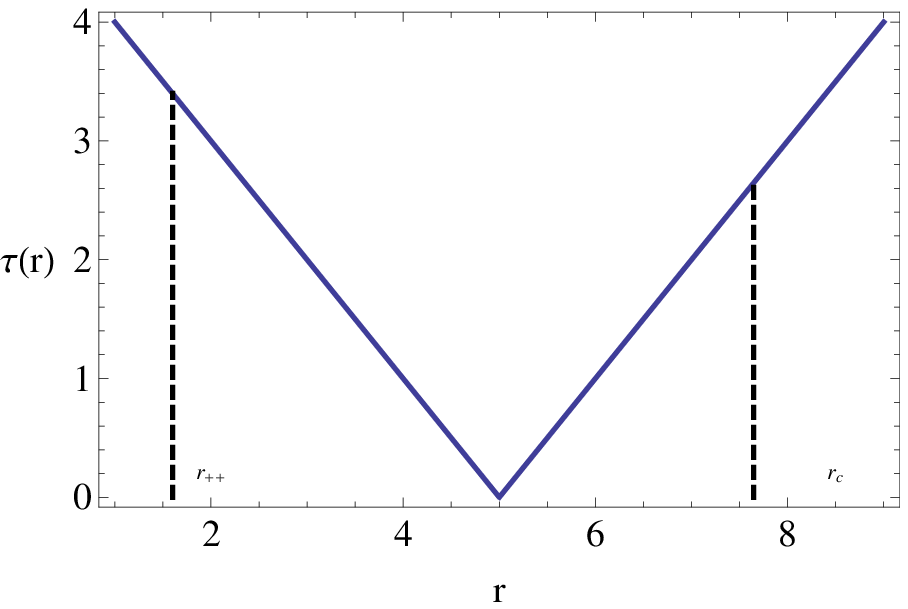}}

\vspace{0.3cm}

 \end{center}

Figure 5. The figure shows  $\tau$ vs $r$ for the black hole with $M=0.96$, $Q=0.96, E=1$ and $\alpha = 0.09$.  Here, $r = 5$ at $ \tau =0$.\\

When the black hole becomes extreme,  the horizons becomes degenerate  with $r_{++}= r_c$. Lets call it $\rho$.  In this case, $f(r)$ is given by the expression,
\be \label{extreme}
f(r) = -\frac{ \alpha (r-r_+) ( r -\rho)^2}{ r^2}
\ee
Hence, for $r > \rho$, $f(r) < 0$. Therefore, $r$ becomes a time coordinate and $t$ becomes a spatial coordinate.  The geodesics equations will be the same as before. The equation for the proper time will be the same and the solutions will be the same as in eq.$\refb{proper}$ with the ``-''  sign in front. The reason to pick the ``-'' sign is due to the fact that for a photon falling towards the horizon from $r > \rho$, the proper time has to be positive.  Once again, the proper time is finite for an in falling photon to reach the horizon. On the other hand, the time $t$ can be obtained by integrating eq.$\refb{dtdr}$ with the function $f(r)$ given in eq.$\refb{extreme}$. The solution is given by,
\be
t =  - \frac{ \rho^2}{ \alpha ( r - \rho) ( r_+ - \rho)} -  \frac{ \left( r_+^2 Log(r - r_+) + ( \rho^2 - 2 r_+ \rho) Log(r - \rho) \right)}{( r_+ - \rho)^2} + constant
\ee
  The integration constant is chosen such that $ t =0$ for $ r = r_0 > \rho$. When   $ r \ra \rho$, the time $t$  reach  $ + \infty$. The ``+'' sign has to be picked from eq.$\refb{dtdr}$ so that  $t$ increases when the photon falls towards the horizons.


\section{ Null geodesics with angular momentum}

In this section, we will study the orbits for photons with non-zero angular momentum. Since $ \dot{r}^2 + V_{eff} = E^2$, the motion of the particles will depend on the values of $E$.  The effective potential for various values of $E$ are plotted in Fig.6. Three different scenarios depending on the values of $E$ are discussed below.

\begin{center}
\scalebox{.9}{\includegraphics{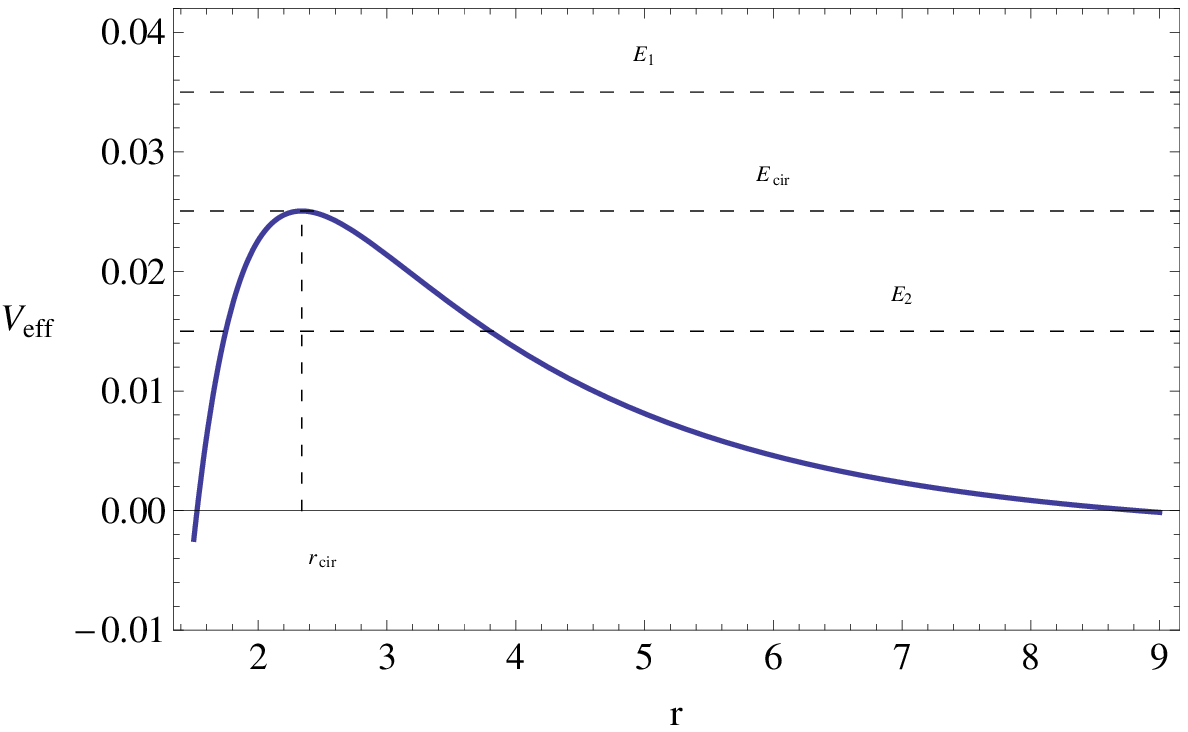}}

\vspace{0.3cm}

 \end{center}

Figure 6. The figure shows  $V_{eff}$ vs $r$ for the black hole with $M=0.96$, $Q=0.96$ and $\alpha=0.09$.\\

{$\bf  E = E_1:$}\\
\noindent
In this case $E_1^2 - V_{eff} > 0$ for all $r$. Hence $\dot{r}^2 >0$ for all $r$ and a photon starting the motion at $r > r_{++}$ will fall into the black hole crossing the horizon.\\

{$\bf E = E_{cir}:$}\\
\noindent
In this case, $\dot{r} =0$ and the orbits are circular.  These are unstable orbits which is evident from the nature of the potential.\\

{$\bf E = E_2:$}\\
\noindent
In this case $\dot{r}^2 >0$ only in two regions. If the initial position is far from the black hole, then the photon will  have a turning point and will not fall into the black hole. If $r_{initial} < r_{cir}$, then the photon will fall into the black hole.


\subsection{ Circular orbits}
From the plot for $V_{eff}$, it is clear that there is an unstable circular orbit. For circular orbits to occur, $ V_{eff} = E_{cir}^2$. At the circular orbits, 

 \begin{equation} \label{dvdr}
  \frac {dV_{eff}}{dr} =0
  \end{equation}{
  which leads to,
  \begin{equation} \label{cube}
  \alpha r^3 - 2 r^2 + 6 M r - 4 Q^2=0
  \end{equation}
 Equation $\refb{cube}$ is a cubic equation which  has three roots. By observing the behavior of the potential by means of graphical presentation, we conclude that the root corresponding to the unstable  circular orbit is the second root given by,
  \begin{equation}
  r_{cir} = 2 \sqrt{ \frac{-\nu_1}{3}} cos ( \frac{\psi_1}{3} - \frac{2 \pi} {3} ) + \frac{ 2}{ 3 \alpha}
  \end{equation}
  where,
  \begin{equation}
  \nu_1 =  \frac{ ( -4 + 18 \alpha M)}{ 3 \alpha^2}
  \end{equation}
  \begin{equation}
  \sigma_1 = \frac{ ( -16 + 108 \alpha M - 108 \alpha^2 Q^2)}{ 27 \alpha^3}
  \end{equation}
  \begin{equation}
  \psi_1 = cos^{-1} \left( \frac{ 3 \sigma_1}{ 2 \nu_1} \sqrt{ \frac{ -3}{ \nu_1}} \right)
  \end{equation}
  In Fig.7,  $r_{cir}$ is plotted with $\alpha$. From the figure, the radius of the circular orbit with the quintessence is larger. In Fig. 8, radius of the circular orbit is plotted as a function of $Q$. When charge increases, $r_{cir}$ decreases.

 \begin{center}
\scalebox{.9}{\includegraphics{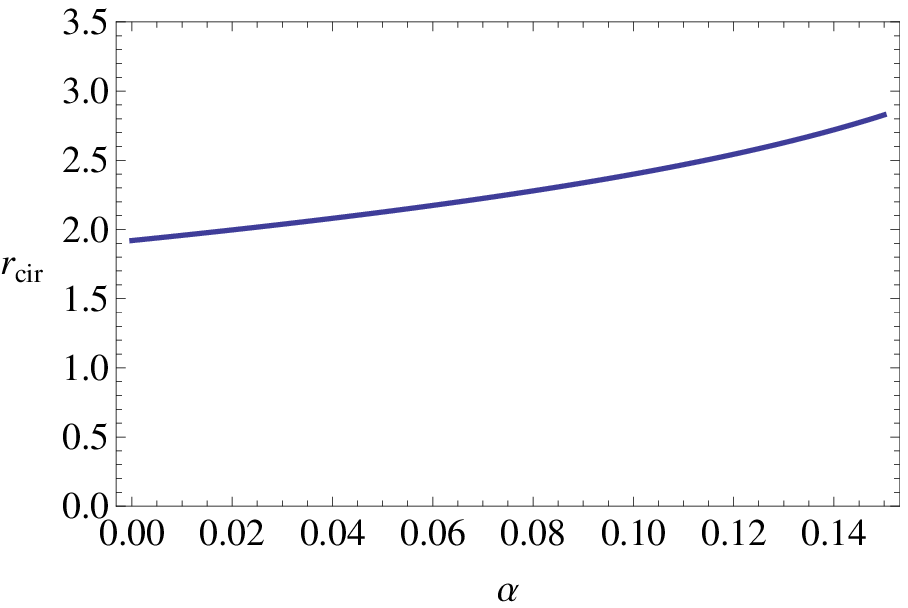}}

\vspace{0.3cm}

 \end{center}

Figure 7. The figure shows  $r_{cir}$ vs $\alpha$ for the black hole with $M=0.96$, and $Q=0.96$.\\

 \begin{center}
\scalebox{.9}{\includegraphics{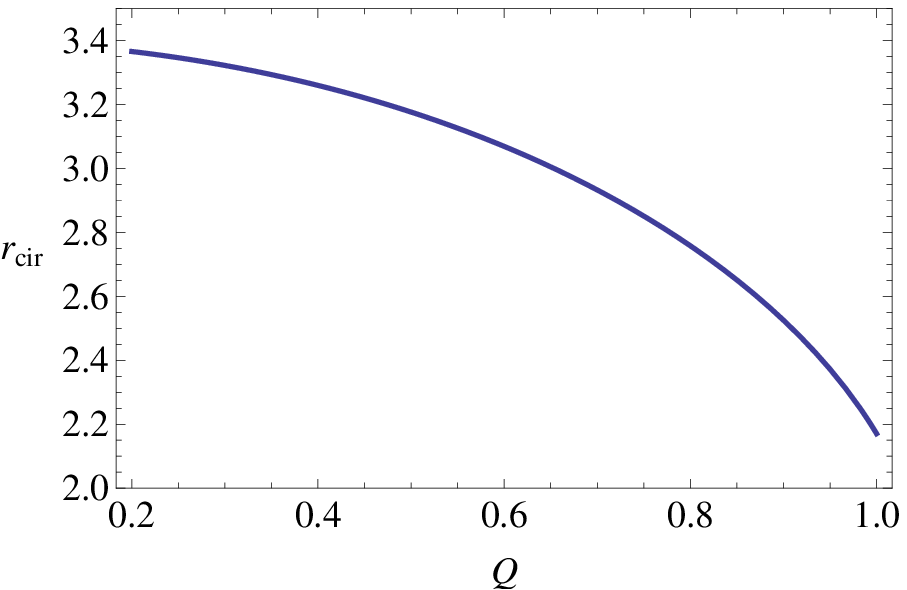}}
\vspace{0.3cm}
 \end{center}

Figure 8. The figure shows  $r_{cir}$ vs $Q$ for the black hole with $M=0.96$, and $\alpha = 0.09$.\\

\noindent
The circular orbit at  $r = r_{cir}$ is called the ``photon sphere'' \cite{vir}.  $r_{cir}$ is independent of $E$ and $L$.  One must note that at the circular orbit, the following condition is also true. 
 \begin{equation} \label{potcir}
V_{eff} = E_{cir}^2
\end{equation}
From eq.$\refb{potcir}$,
\begin{equation}
\frac{E_{cir}^2} {L_{cir}^2}=  \frac{ f( r_{cir}) } {r_{cir}^2} \Rightarrow 
\frac{ 1}{D^2_{cir}} = \frac{ Q^2}{ r_{cir}^4}  - \frac{ 2 M}{ r_{cir}^3} + \frac{1}{r_{cir}^2} - \frac{ \alpha}{r_{cir}} 
\ee
$D_{cir} $ is the critical impact parameter given by $\frac{ L_{cir}} { E_{cir}}$. In Fig.9,  $D_{cir}$ is plotted as a function of $\alpha$. $D_{cir}$ increases with $\alpha$. Hence the black  hole surrounded by dark energy requires higher impact parameter to orbit in a circle around the black hole. Also, for fixed value of $L_{cir}$, $E_{cir}$ decreases when $\alpha$ increases. Hence, photons needs less energy to go in circular orbits for higher $\alpha$. As a consequence, the value of $E_1$ and $E_2$ also decrease for higher $\alpha$.

\begin{center}
\scalebox{.9}{\includegraphics{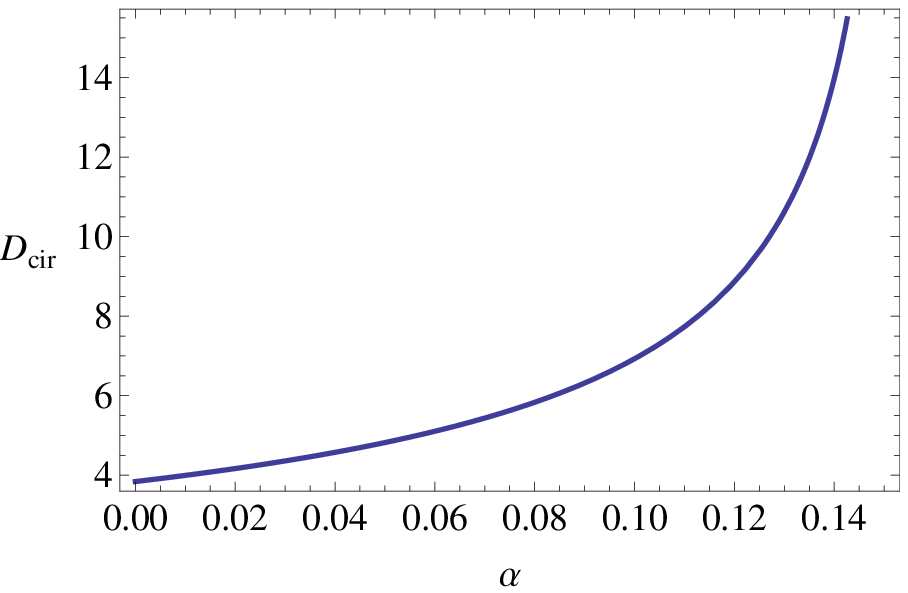}}

\vspace{0.3cm}

 \end{center}

Figure 9. The figure shows  $D_{cir}$ vs $\alpha$ for  a black hole with $M=0.96$, and $Q=0.96$.\\


\subsection{ Geodesics with the variable $u$ }

One of the main goals of this paper is to obtain analytical expressions for the trajectories of the photon orbits. Analytical solutions offer high  accuracy compared with numerical calculations. High accuarcy is needed if one like to make observations and make comparisons with theoretical predictions. In order to obtain analytical observations, we would like to make change of variable, $ u = \frac{1}{r}$ to study the orbits. First, one can rewrite the eq.$\refb{drdphi}$ as,
\begin{equation} \label{dudphi}
\left(\frac{ d u}{ d \phi} \right)^2 = g(u)
\end{equation}
where,
\begin{equation}
g(u) =  2 M u^3  -   u^2  + \alpha u  + \frac{ E^2}{L^2} - Q^2 u^4
\end{equation}
The geometry of the geodesics will depend on the roots of the function $g(u)$. Notice that when $ u \rightarrow \pm \infty$, $g(u) \rightarrow - \infty$.
Also, when $u \rightarrow 0$, $g(u) \rightarrow  \frac{ E^2}{L^2}$. Therefore, $g(u)$ always have two real roots and one of them is negative and the other a positive one. Since $g(u)$ is a polynomial of fourth order, it is possible for $g(u)$ to have two more real roots or a complex conjugate pair. It is also possible for $g(u)$ to have degenerate roots. All these possibilities are shown in Fig.10.

\begin{center}
\scalebox{.9}{\includegraphics{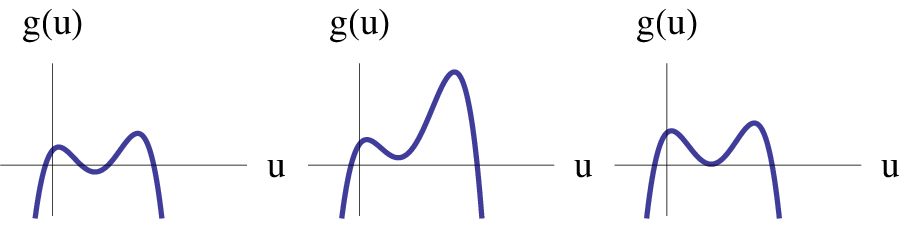}}\\
\vspace{0.2cm}
\end{center}
Figure 10. The graph shows the function $g(u)$ as a function of $u$.\\

From eq.$\refb{dudphi}$,

\begin{equation}
\left( \frac{ du}{d \phi} \right) = \pm \sqrt{ g(u) }
\end{equation}
where $g(u)$ is written as,
\begin{equation}
g(u) = - Q^2 ( u - u_1) ( u - u_2) ( u - u_3) ( u - u_4)
\end{equation}
Here, the ``+'' sign will be chosen without lose of generality.  When the above equation is integrated,  a relation between $u$ and $\phi$  in terms of Jacobi-elliptic integral $\mathcal{F}( \xi, y)$ is obtained as,
\begin{equation}
\phi = \frac { 2}{Q}  \frac{ \mathcal{F} ( \xi, y)}{ \sqrt{ ( u_2 - u_3) ( u_1 - u_4)}} + constant
\end{equation}
Here,
\begin{equation}
sin \xi  = \sqrt{ \frac{ ( u - u_2)( u_1 - u_4)}{ ( u  - u_1) ( u_2 - u_4)}}
\end{equation}
\begin{equation}
y = \frac{( u_1- u_3) ( u_2 - u_4) }{ ( u_2 - u_3)(u_1 - u_4)}
\end{equation}
\be
\mathcal{F}(\xi, y) =  \int^{\xi}_0 \frac{ d \lambda}{ \sqrt{ 1 - y sin^2(\lambda)}}
\ee
The constant in the equation for $\phi$ is an integration constant and could be complex depending on the values of the roots. However, the final result for $\phi$ will be real.

\subsection{ Orbits for various values of  energy $E$}
Depending on the values of $E, M, Q, \alpha$ and $L$, the root structure of $g(u)$ varies. In the following, we will highlight separate cases.\\

{$\bf  E > E_{cir}$}:\\
\noindent

In this case, the plot of $g(u)$ is given in Fig.11. The corresponding orbits for two values of energy $E$ is given in the Fig.12. The one with less energy falls into the black hole earlier than the one with more energy as expected.

\begin{center}
\scalebox{.9}{\includegraphics{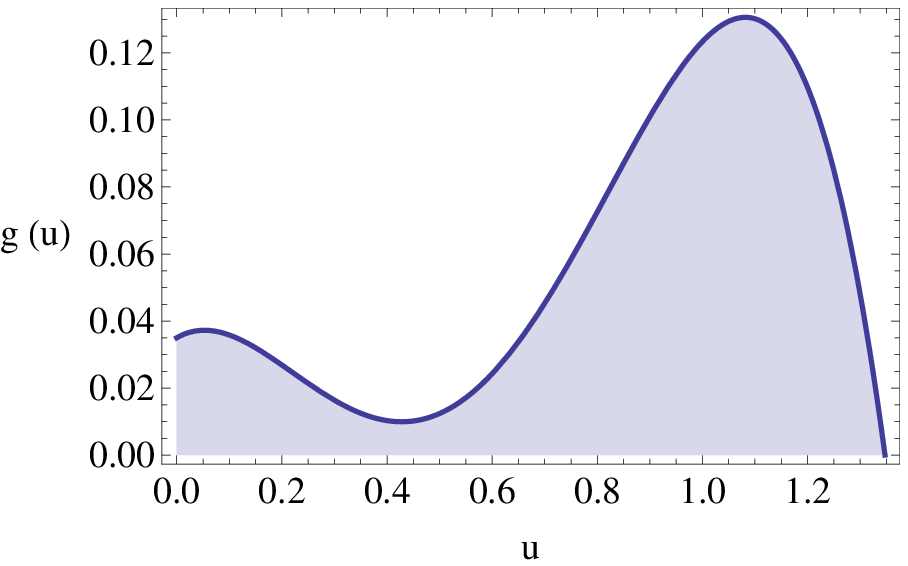}}\\
\vspace{0.2cm}
\end{center}
Figure 11. The graph shows the function $g(u)$ as a function of $u$. Here, $ M = 0.96, Q = 0.96,  L=1, \alpha = 0.09$ and $ E = 0.187$. 

\begin{center}
\scalebox{.9}{\includegraphics{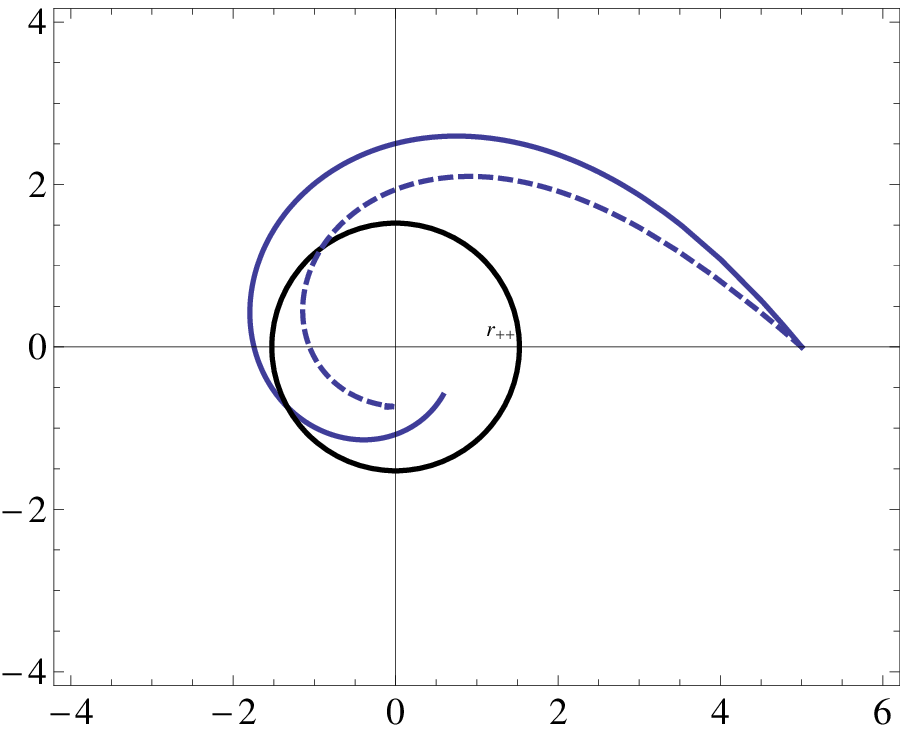}}\\
\vspace{0.2cm}
\end{center}
Figure 12. The graph shows the polar plot of $\phi$ of the particle with energies  $ E = 0.187$(thick) and $E=0.247$(dashed). The black hole has $ M = 0.96, Q = 0.96, L=1$,  and $ \alpha = 0.09$. The circle is the black hole horizon.\\

{$\bf  E = E_{cir}$}:\\
\noi

In this case, the plot of $g(u)$ is given in Fig.13. The corresponding orbit is given in Fig.14. There are two orbits with one starting inside the circular orbit and the other starting out side the circular orbit. Since they both have the same critical energy, $E_{cir}$, they both reach the unstable circular orbit at $r = r_{cir}$.

\begin{center}
\scalebox{.9}{\includegraphics{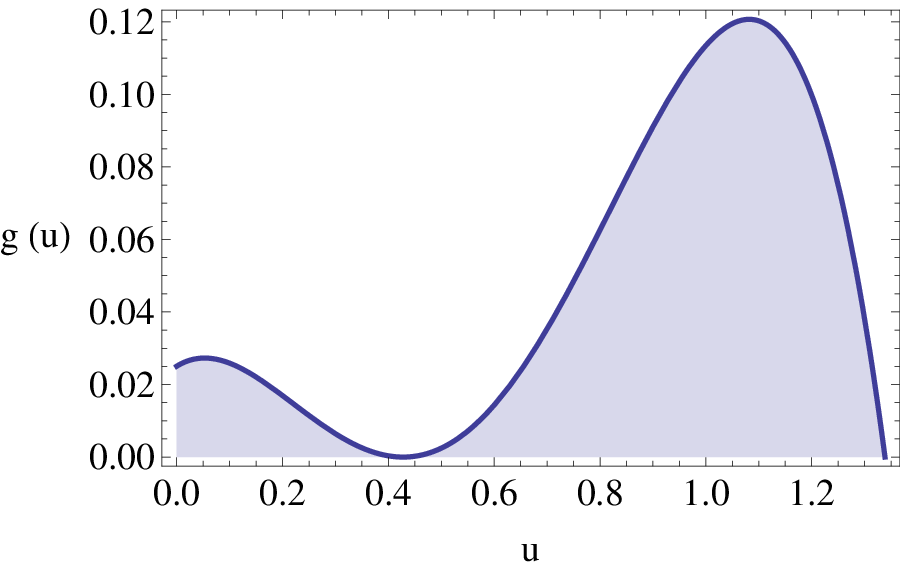}}\\
\vspace{0.2cm}
\end{center}
Figure 13. The graph shows the function $g(u)$ as a function of $u$. Here, $ M = 0.96, Q = 0.96, L=1,\alpha = 0.09$ and $ E = 0.158$\\

\begin{center}
\scalebox{.9}{\includegraphics{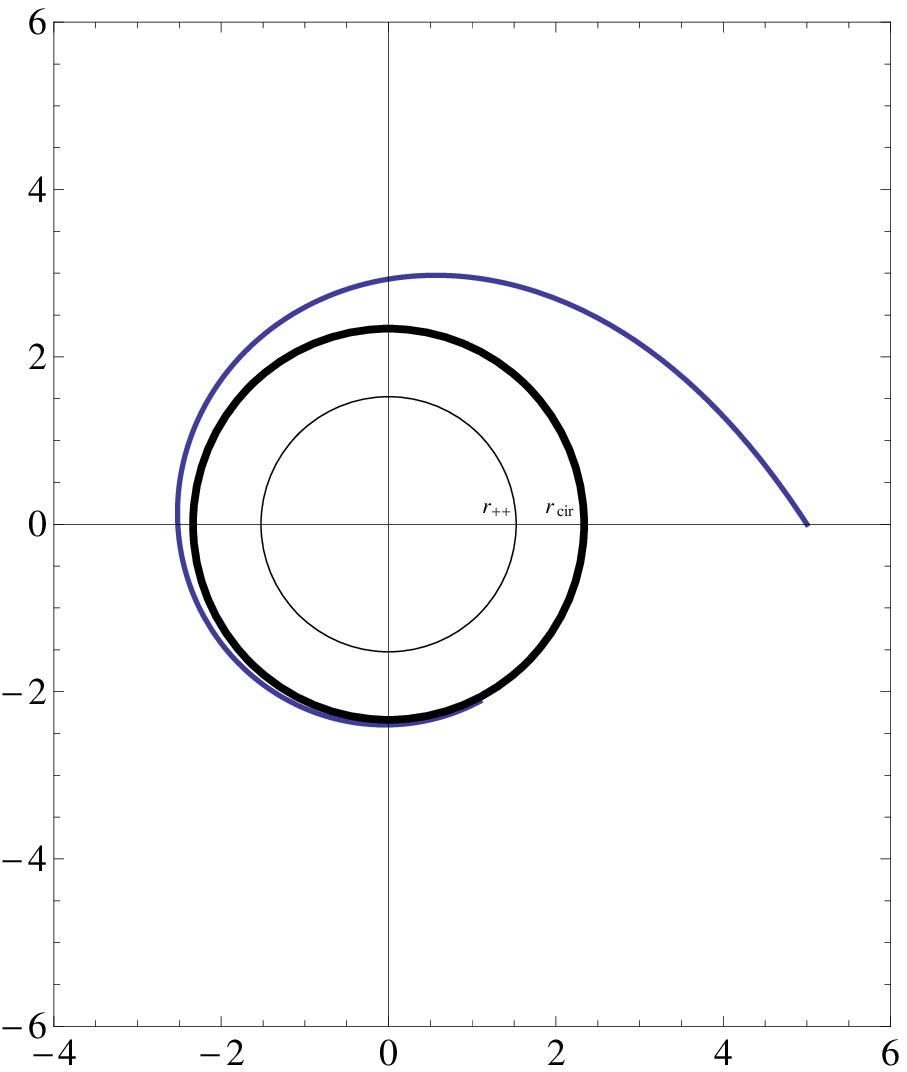}}\\
\vspace{0.2cm}
\end{center}
Figure 14. The graph shows the polar plot of $\phi$ of the particle with $ E = 0.158$. The black hole has $ M = 0.96, Q = 0.96, L=1$ and  $\alpha = 0.09$. The trajectory from out side approaches the black hole and merge to the circular orbit at $r = r_{cir}$. The trajectory from inside given by the dashed curve approaches the circular orbit from inside and merge the circular orbit at $r = r_{cir}$ as expected.\\

{$\bf  E < E_{cir}$}:\\

In this case, the plot of $g(u)$ is given in Fig.15. The corresponding orbit is given in Fig.16. There are two orbits in the Fig.24. One with low energy (thick curve) and the one with high energy(dashed curve). The one with the high energy do not bend as much as the one with lower energy in its orbit as expected.

\begin{center}
\scalebox{.9}{\includegraphics{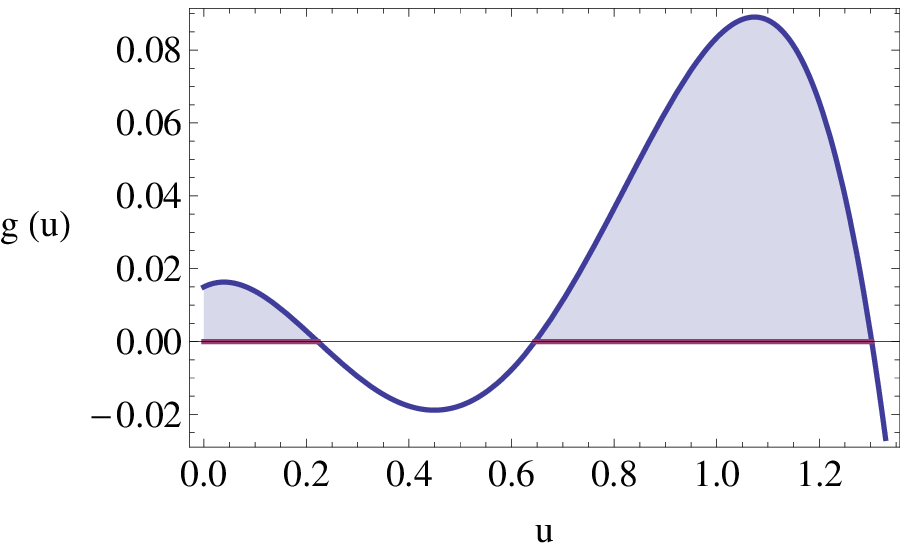}}\\
\vspace{0.2cm}
\end{center}
Figure 15. The graph shows the function $g(u)$ as a function of $u$. Here, $ M = 0.96, Q = 0.96, L =1, \alpha = 0.07$ and $ E = 0.122$\\

\begin{center}
\scalebox{.9}{\includegraphics{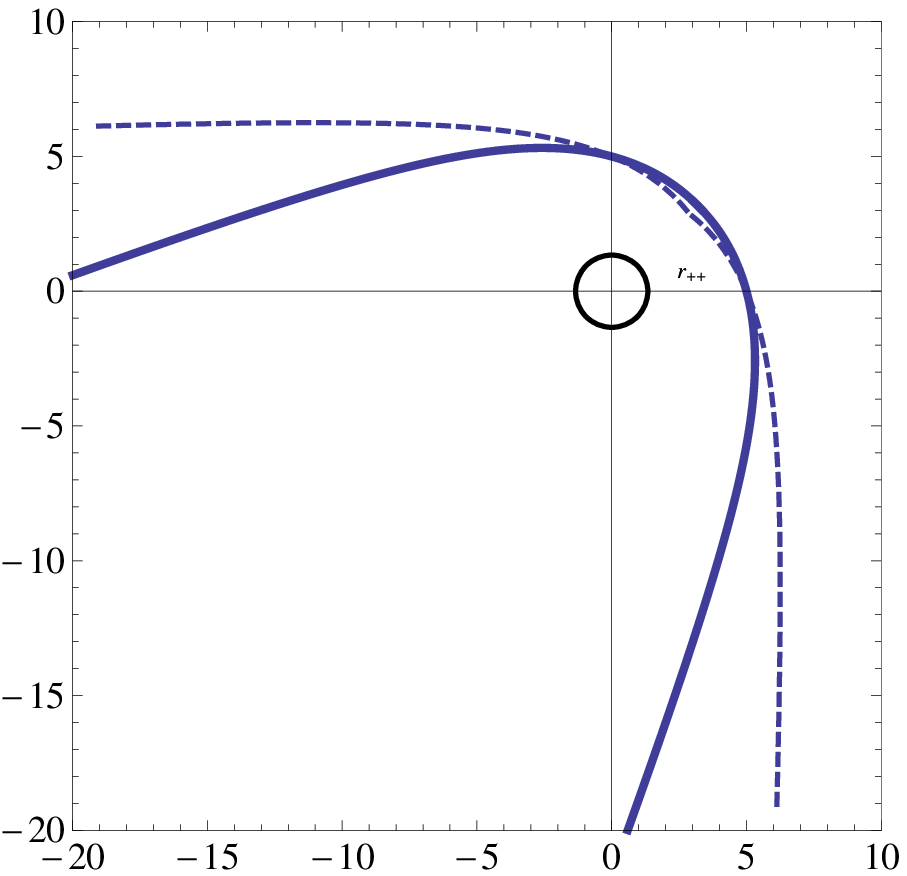}}\\
\vspace{0.2cm}
\end{center}
Figure 16. The graph shows the polar plots of $\phi$ of the particle with energies $ E = 0.122$ (thick)  and $E=0.138$(dashed). The black hole has $ M = 0.96, Q = 0.96,L=1,  \alpha = 0.07$\\


\section{ Bending of light around the black hole}

When photons travel around a black hole, the light bends. This deflection of light is one of the few observational tools available to study the geometry around a black hole or an object with strong gravitational field. There are many works in the literature with focus on bending of light around black holes. We will mention only a few that may be related to the present work. The deflection of light and the gravitational frequency shift of the Schwarzschild black hole surrounded by the quintessence  was studied by Liu et.el \cite{liu2}. Light bending as  a probe of the dark energy  was presented by Finelli et.al. in \cite{fine}.  Bending of light in conformal gravity was presented in \cite{sul}. Bending of light and the motion of particles in the background of the Schwarzschild-de Sitter black hole is well studied by many authors: bending of light is studied  by Ishak et.al.\cite{ishak}, and gravitational lensing has been studied by Sch$\ddot{u}$cker \cite{schu} and Sereno \cite{sereno}. A comprehensive study of motion of particles around the Schwarzschild-de Sitter black hole was done by Struchlik \cite{struch}. Properties of the motion of massive particles and photons around the Reissner-Nordstrom-de Sitter black holes was studied by Stuchlik and Hledik in \cite{stu2}.

The conventional approach of calculating the angle of deflection gives it as \cite{wein},

\be \label{ben}
 \bigtriangleup \varphi = 2 | \phi( \infty) - \phi( r_o) | - \pi
 \ee
where $r_o$ is the closest approach of the photon when it travels around the black hole. However, this approach works only for  asymptotically flat space-times. In the current work, the space-time is asymptotically non-flat and $ r \ra \infty$ does not make sense. To remedy this situation, Rindler and Ishak developed a method to find the bending angle in asymptotically non-flat space-times \cite{ishak2}.

In particular, they applied this approach to find the bending angle of Schwarzschild-de Sitter black hole to find the contribution of the cosmological constant on the bending angle. Here, we will follow their approach to compute the bending angle for the charged black hole with the quintessence.


\subsection{Rindler Ishak method to find the angle of deflection}

Rindler-Ishak method  is based on the Figure. 17. They defined the general bending angle as $ \epsilon = \psi - \phi$. Hence, the total bending angle was defined as  $2 \epsilon$ when $ \phi =0$, which gives $ 2 \epsilon_0 = 2 \psi_0$. The angle $\psi$ is given by,

\be \label{bend}
tan \psi = \frac{ \sqrt{ f(r)} r}{ |A|}
\ee
where
\be
A = \frac{ dr}{ d\phi}
\ee
For more details on the derivation of equation $\refb{bend}$, the reader is referred to the paper by Rindler and Ishak \cite{ishak2}.

\begin{center}
\scalebox{.9}{\includegraphics{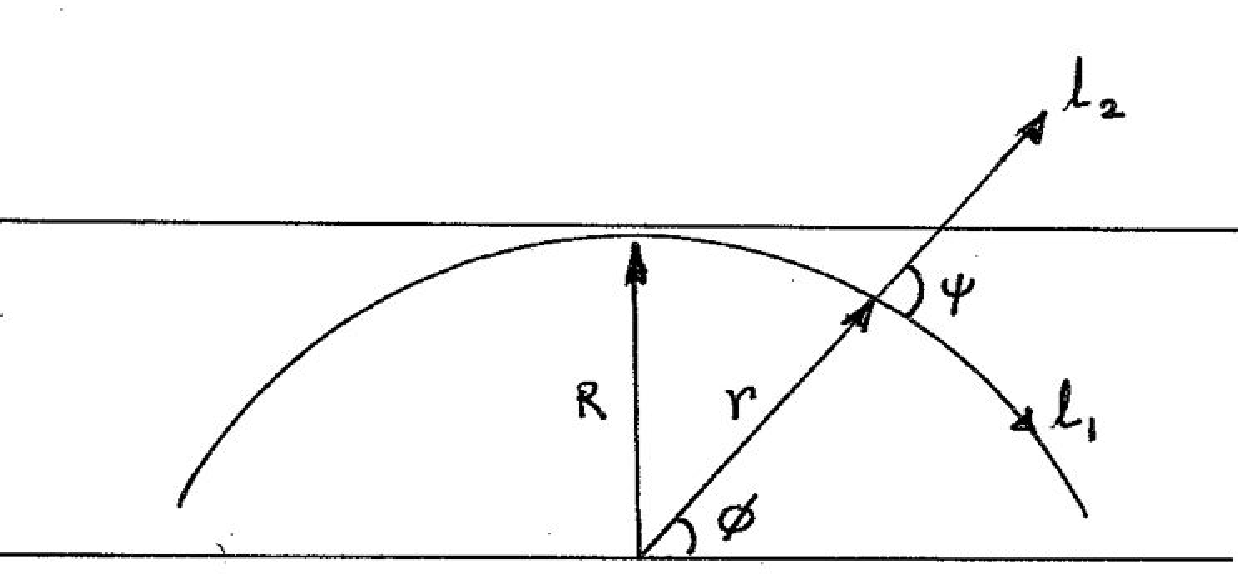}}\\
\vspace{0.2cm}
\end{center}
Figure 17. The figure shows a light ray bending around the black hole.


\subsection{ Perturbative approach to find $ u = \frac{1}{r}$ as a function of $\phi$.}

In this section we will present how we found a solution for $u$ in terms of $\phi$ using a perturbative approach. From equation$\refb{dudphi}$,  which corresponds to the photon path around the black hole, one can obtain a second order differential equation,
\be \label{orbit}
\frac{ d^2u}{ d \phi^2} + u -  \frac{ \alpha}{2} = 3 M u^2 - 2 Q^2 u^3
\ee
This is a non-linear differential equation which can be solved perturbatively in powers of $M$ and $Q^2$. Let the solution for $ M =0, Q=0$ be $u_0$ and a small perturbation be $u_1$. Hence to first order, $ u = u_0 + u_1$. By substituting this to the equation $\refb{orbit}$, we get two equation,

\be  \label{u0}
\frac{ d^2u_0}{ d \phi^2} + u_0 -  \frac{ \alpha}{2} = 0
\ee
and,
\be \label{u1}
\frac{ d^2u_1}{ d \phi^2} + u_1 = 3 M u_0^2 - 2 Q^2 u_0^3
\ee
First, let us solve the zeroth order equation given by,
\be  \label{u0}
\frac{ d^2u_0}{ d \phi^2} + u_0 -  \frac{ \alpha}{2} = 0
\ee
Let $ \bar{u} = u - \alpha/2$.
Then, equation$\refb{u0}$ simplifies into,
\be  \label{newu0}
\frac{ d^2 \bar{u}_0}{ d \phi^2} + \bar{u}_0  = 0
\ee
The solution to equation$\refb{newu0}$ is,
\be
\bar{u} = \frac{ cos \phi}{R}
\ee
$R$ is a constant which appears in Figure.17. Hence the solution for $u_0$ is,
\be \label{solu0}
u_0 =  \frac{ cos \phi}{R}  + \frac{\alpha}{2}
\ee
Now, the solution  for $u_0$   can be substituted to the right hand of the first order equation$\refb{u1}$ leading to,

\be \label{newu1}
\frac{ d^2u_1}{ d \phi^2} + u_1 = 3 M \left(\frac{ cos \phi}{R}  + \frac{\alpha}{2}\right)^ 2 - 2 Q^2 \left (\frac{ cos \phi}{R}  + \frac{\alpha}{2}\right)^3
\ee
The final solution   $u = u_0 + u_1$. Replace   $\phi$ with $ (\pi/2 - \phi)$ to reflect the angles in Figure 17. Then, the  solution $u$  is given by,
\be
u =  \frac{\alpha}{2} + \frac{ 3 M \alpha^2}{4} - \frac{ Q^2 \alpha^3}{ 4}  + \frac{ B_1}{R} + \frac{ B_2}{R^2} + \frac{ B_3}{R^3}
\ee
where,
$$
B_1 =  sin \phi + M \left( \frac{ 3 \alpha \pi}{4} cos \phi - \frac{3}{2} \phi \alpha cos \phi + \frac{3}{2} \alpha sin \phi \right) +
$$
\be
Q^2 \left( - \frac{3}{8} \pi \alpha^2 cos \phi + \frac{3}{4} \phi \alpha^2 cos \phi - \frac{3}{4} \alpha^2 sin \phi \right)
\ee

\be
B_2 = M \left( \frac{3}{2} + \frac{1}{2} cos 2 \phi \right) + Q^2 \left( - \frac{ 3 \alpha}{2} - \frac{ \alpha}{2} cos 2 \phi \right)
\ee

\be
B_3 = Q^2 \left( - \frac{ 3}{8} \pi cos \phi + \frac{3}{4} \phi cos \phi -\frac{9}{16} sin \phi - \frac{1}{16} sin 3 \phi \right)
\ee
When $\alpha \ra 0$ and $ Q \ra 0$, one obtain,
\be
u = \frac{sin \phi}{R} + \frac{ M}{2 R^2} ( 3 + cos 2 \phi)
\ee
which is the $u$ value obtained by  Rindler and Ishak \cite{ishak2} for the Schwarzschild-de Sitter black hole.

Now in order to find the bending angle, one can us the fact that $r = \frac{1}{u}$ and let $\phi =0$ to obtain
\be
r(\phi=0) =  r_z =  \frac{ 8 R^3}{X}
\ee
and
\be
A = \frac{dr}{d \phi} ( \phi=0) = \frac{ -64 R^5}{X^2}
\ee
where,
\be
X = - 3 \pi \left( - 2 MR^2 \alpha + Q^2 ( 1 + R^2 \alpha^2)\right) + 2 R \left( M ( 8 + 3 R^2 \alpha^2) - \alpha( - 2 R^2 + Q^2 ( 8 + R^2 \alpha^2)) \right)
\ee
Now, one can utilize the equation $\refb{bend}$ to find the bending angle $\psi_0$. 

\be \label{bendactual}
tan \psi = \sqrt{ f(r_{\phi=0})}  \frac{ |X|}{ 8 R^2}
\ee
We have taken the absolute value of $X$ since one has to have the absolute value of $A$ in the expression. Now, we will expand $\sqrt{f}$ for small values of $M/R$, $Q^2/R$, $ \alpha R$  and substitute $r_z$ for $r$ leading to,
\be
\sqrt{f} \approx 1 - \frac{ M} { r_z} + \frac{ Q^2}{ 2 r_z^2} - \frac{ \alpha r_z}{ 2}
\ee
Also, we will assume that the angle $\psi$ is small. Hence $tan \psi$ $\approx \psi$. Hence the angle $\psi$ at $ \phi=0$, which is $\psi_0$ is,
\be 
\psi_o  \approx \left(1 - \frac{ M} { r_z} + \frac{ Q^2}{ 2 r_z^2} - \frac{ \alpha r_z}{ 2} \right) \frac{ |X|}{ 8 R^2}
\ee

By  substituting $X$ and $r_z$ into the above expression and expanding for small $M/R$, $Q^2/R^2$, and $ \alpha R$, we obtain $\psi_0$. Therefore, the bending angle, $2 \psi_0$  is given by,
\be \label{approxbending}
2 \psi_0 = \frac{ 4 M}{R} + \frac{ 3 \pi M \alpha}{2} - \frac{ 4 M^2 \alpha}{R} - \frac{ 3 \pi Q^2}{ 4 R^2} - \frac{ 4 Q^2 \alpha}{ R} + \frac{ 3 \pi M^2 Q^2 }{2 R^4}
\ee
Notice that  the first term is the one for Schwarzschild black hole.  Other terms are corrections due to $\alpha$ and $Q^2$. We have omitted higher order terms here. We have plotted the bending angle vs $\alpha$ in Fig.18.  For the given parameters, the bending angle increases with $\alpha$. However, taking a closer look at the eq.$\refb{approxbending}$, it is evident that the behavior of the bending angle depend on if the xpression,
\be
 \frac{ 3 \pi M }{2} - \frac{ 4 M^2 }{R}  - \frac{ 4 Q^2}{ R}
\ee
is positive or negative. 

\begin{center}
\scalebox{.9}{\includegraphics{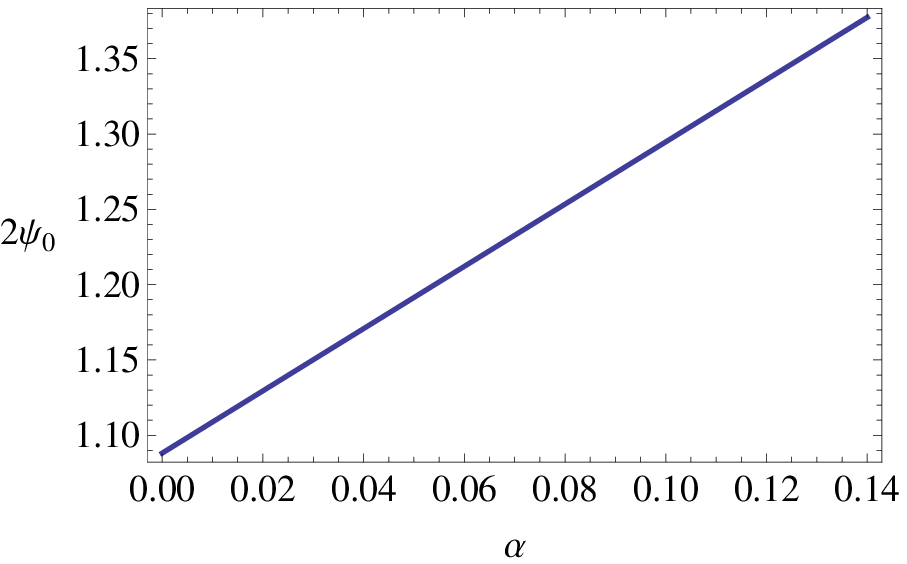}}\\
\vspace{0.2cm}
\end{center}

Figure 18. The figure shows  the bending angle $2 \psi_0$ vs $\alpha$. Here, $M = 0.96, Q = 0.96$ and $ R = 3$.

Another work which has followed Rindler and  Ishak method to compute bending angle is given in \cite{carlo}.


\section{ Conclusion}

The objective of this paper is to study a black hole surrounded by a scalar field called ``quintessence.'' The quintessence field is one of possible sources for dark energy considered in research currently. Our main goal in this paper was to study null trajectories around a charged black hole surrounded by the quintessence derived by Kiselev \cite{kiselev}.

First we presented the structure of horizons where it was shown that the black hole can have three, two or one horizons. Similar to de-Sitter black holes,  the one with the quintessence also has a cosmological horizon. 

To understand  the trajectories,  we did  a detailed study of  null geodesics around the charged black hole surrounded by  quintessence.  Both  radial and the trajectories with angular momentum are studied.  The exact solution to the geodesics equation  with angular momentum is obtained in terms of Jacobi-elliptic  integrals. The orbits for various values of  the energy  are presented. The circular orbits were studied in detail which were unstable. The radius of the circular orbits were larger for large $\alpha$ and small for large $Q$.

As an application of the photon trajectories, we have also studied the bending angle of light. We have used a method developed by Rindler and Ishak \cite{ishak2}. First  we used a perturbative approach to obtain the solutions for the trajectory. Then we used expansions around small $M/R$, $Q^2/R^2$ and $\alpha R$, to obtain  the bending angle.

In this paper we have focused on the photon motion for black holes with non-degenerate horizons. The geodesics for the extreme black holes and the global structure of the extreme black holes would be an interesting avenue to study. For example, the structure of the extreme Schwarzschild-de Sitter space-time is studied by Podolsky in \cite{pod}. We are planning on reporting on  the extreme charged black hole with  quintessence in the future.

It would also be interesting to  do further analysis of strong gravitational lensing along the lines of the work done by Virbhadra et.al \cite{vir2}.



\end{document}